\documentstyle[12pt,epsfig]{article}

\newlength{\dinwidth}
\newlength{\dinmargin}
\makeatletter
\def\gappr{\mathpalette\under@rel{>\approx}}
\def\lappr{\mathpalette\under@rel{<\approx}}
\def\gsim{\mathpalette\under@rel{>\sim}}
\def\lsim{\mathpalette\under@rel{<\sim}}
\def\under@rel#1#2{\under@@rel#1#2}
\def\under@@rel#1#2#3{\mathrel{\vcenter{\hbox{$%
  \lower3.8pt\hbox{$#1#2$}\atop{\raise1.8pt\hbox{$#1#3$}}%
  $}}}}
\makeatother

\newcounter{subequation}[equation]

\makeatletter
\expandafter\let\expandafter\reset@font\csname reset@font\endcsname
  {\endeqnarray\stepcounter{equation}}
\makeatother

\newcounter{statement}
\newenvironment{statement}[4]
  {\par\refstepcounter{statement}
    \noindent#1#2 \arabic{statement} #4\unskip: #3}{\par\vspace{2mm}}

\newenvironment{statement*}[4]
  {\par\noindent#1#2 #4\unskip: #3}{\par\vspace{2mm}}

\begin{document}
\hbox to\hsize{%
\hfil
 \vbox{%
        \hbox{MPI-PhT/??}%
        \hbox{May, 1996}%
        \hbox{gr-qc/9605053}%
        }}

\vspace{1cm}
\begin{center}
\LARGE\bf
Solitons of the Einstein-Yang-Mills Theory\footnote{To appear in the
proceedings of the {\it Pacific Conference on Gravitation and Cosmology}, Seoul, Feb. 1996}

\vskip5mm
\large 
Dieter Maison\footnote{e-mail: dim@mppmu.mpg.de}

\vspace{3mm}
\small\sl
Max-Planck-Institut f\"ur Physik\\
--- Werner Heisenberg Institut ---\\
F\"ohringer Ring 6\\
80805 Munich (Fed. Rep. Germany)

\end{center}
\vspace{10mm}
\begingroup \addtolength{\leftskip}{1cm} \addtolength{\rightskip}{1cm}
\begin{center}\large\bf Abstract\end{center}
\vspace{3mm}\noindent
Subject of this talk is an overview of  
results on self-gravitating solitons of the classical Yang-Mills-Higgs theory.
One finds essentially two classes of solitons, one of them   
corresponding to the magnetic
monopoles the other one to the sphalerons of flat space. 
The coupling to the gravitational field leads to new
features absent in flat space. These 
are the gravitational instability of these solitons at the Planck scale and
the existence of
black holes with `non-abelin hair'' in addition to the regular solutions.
\endgroup
\vspace{1cm}
\newpage
\section{Introduction}

My talk is an overview of results on self-gravitating solitons of the classical
Yang-Mills-Higgs (YMH) theory. It is based on analytical and
numerical results obtained in collaboration with P.~Breitenlohner and 
P.~Forg\'acs \cite{BFM} . Many other people, who have
contributed in establishing our present understanding of this subject
will be mentioned in due course.

Let me start by specifying, what precisely I understand under solitons,
since this concept is used with various different meanings in the literature.
I will adhere to a rather liberal use of this concept, denoting by it any
particle-like solution of a non-linear field theory.
Particle-like solutions are localized, time-independent 
solutions of finite total energy (mass) with some stability against
perturbations. A typical, maybe the best known example of relativistic
solitons are the non-abelian `t~Hooft-Polyakov monopoles.
As a genuine non-linear structure 
they play an important role in the non-perturbative
aspects of the YMH theory. For a long time it was
believed, that also Einstein's theory of General Relativity had such smooth
solitonic solutions - called geons. However, as was shown by Lichnerowicz
\cite{Lichner},
neither Einstein's theory in vacuum nor the Einstein-Maxwell 
theory give rise to regular particle-like solutions. 
Nevertheless there are alternative
candidates  - the Schwarzschild resp. Reissner-Nord\-str{\o}m (RN) black
holes. Although they suffer from a physical singularity at their center,
this singularity is hidden from the observer behind an event horizon. Black
holes
have finite mass and behave in many ways like genuine particles. In fact,
they may well be considered as ``renormalized'' point particles, dressed
with their gravitational self-field \cite{Damour}.

Regular particle-like solutions were found for models involving 
gravitating complex scalar fields
(``Boson stars''), but they have rather the properties of exotic cosmic
objects than those of particles \cite{Colpi}. 
Later Lichnerowicz's No-Go-Theorem could be generalized 
from the Einstein-Maxwell theory to Kaluza-Klein
models and supergravities \cite{BGM}, leading to the belief that no
smooth solitons can be found for self-gravitating gauge theories.
It is clear that this result cannot apply to self-gravitating versions of
flat-space solitons like 't~Hooft-Polyakov monopoles, first studied by
van~Nieuwenhuizen, Wilkinson and Perry \cite{Perry}. 
However for the YM theory without a
Higgs field, which has no solitons in flat space,
it came as a surprise to many, when Bartnik and McKinnon (BM) \cite{Bart} 
discovered a family
of regular, localized (finite mass) solutions of
the gravitating theory.
Although their discovery was based only on ``numerical evidence''
a rigorous existence proof was found subsequently
(with due delay!) \cite{Yau}.
When it was found that they are unstable these ``particles'' 
lost some of their glamour. It was realized that they had much in common
with the ``sphalerons'' of the YMH theory \cite{Wald}.
 The latter are solutions with a
Higgs doublet (for the gauge group $SU(2)$) in contrast to the monopole
obtained with a triplet.
Varying the strength of the gravitational coupling 
of the gravitating version of the flat-space sphaleron one obtains
a one-parameter family
of solutions interpolating between the flat solution and the (first) 
BM solution \cite{Greene}.
This strongly suggests to interprete the BM solutions as gravitationally
bound counter-parts of the Higgs-bound flat sphaleron. Indeed,
still other sphalerons can be obtained replacing the gravitational field 
by a dilaton \cite{LM}.

Besides the gravitating sphalerons I will also discuss the effects of the
gravitational self-interaction on the non-abelian monopoles. As to be
expected they develop a gravitational instability for sufficiently strong
gravitational self-force. Contrary to naive expectation
(and to claims in the literature) the static
monopoles do however not simply turn into black holes
as the strength of the gravitational
coupling is increased to its critical value. 
For values close to the critical strength of the gravitational
coupling the space-like hypersurfaces of these solutions develop two
distinct regions separated by a long throat. The inner part tends to
a kind of `cosmological''
solution representing a closed asymptotically Robinson-Bertotti universe,
whereas the outer one becomes the exterior part of the extremal RN black
hole.

As already mentioned before there is a second type of soliton
in General Relativity - the black hole. Taking the YM resp.\ YMH model as
the matter part one finds a rich spectrum of static black hole solutions.
This is to be contrasted with Einstein's theory in vacuum resp.\ with the
Einstein-Maxwell theory, where according to a theorem of Israel \cite{Israel} 
the Schwarzschild resp.\ RN solution are the only static black holes.
In the EYM theory one finds
not only the embedding of the abelian RN black hole, but in
addition there are genuinely non-abelian (``coloured'') black holes.
Their co-existence gives rise to an interesting violation of the 
``No-Hair-Conjecture'', since they carry the same (magnetic) charge
\cite{Bizon}.
 
\section{Yang-Mills-Higgs in Flat Space}\label{flat}
\begin{figure}
\hbox to\hsize{
  \epsfig{file=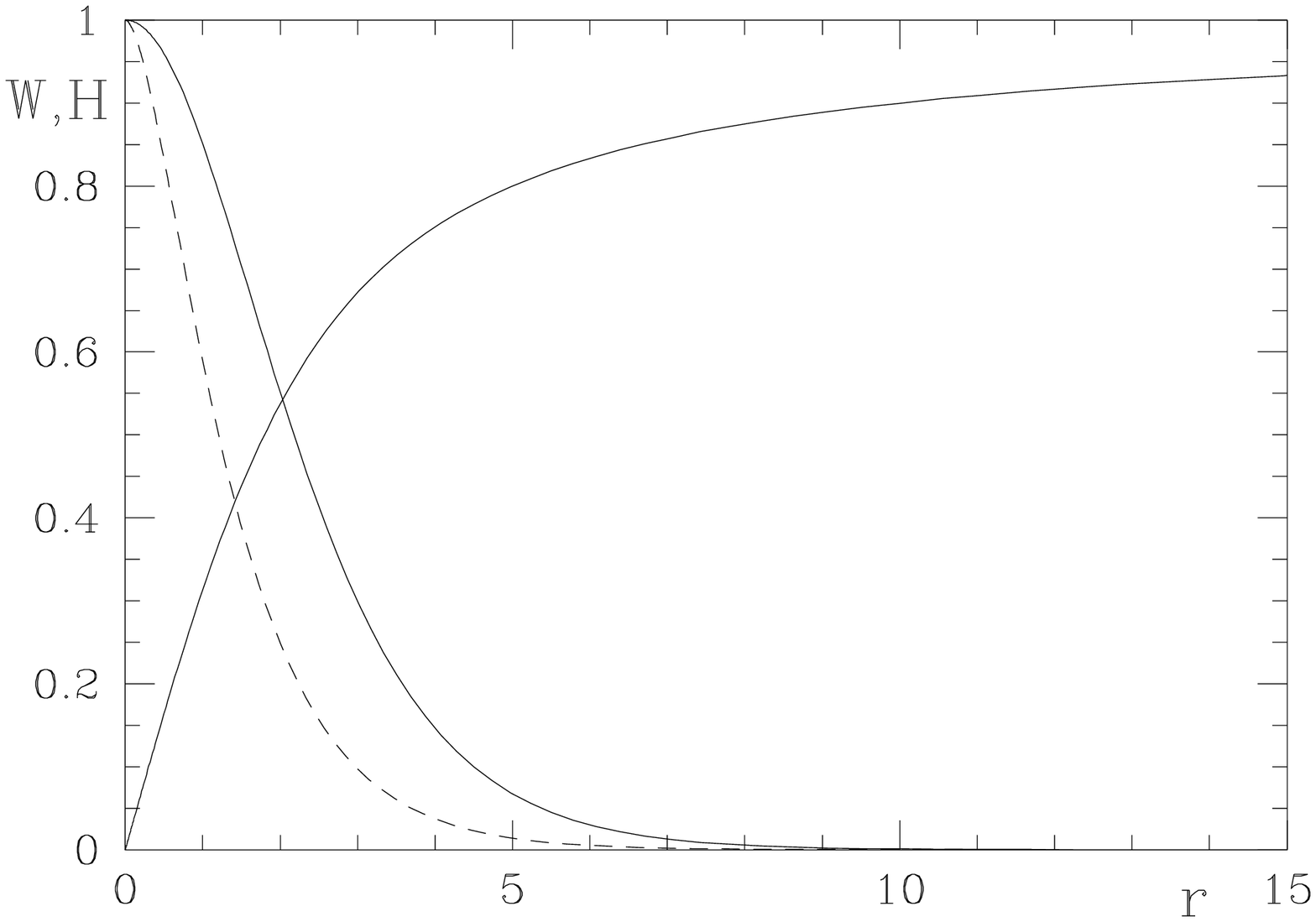,width=0.48\hsize,%
      bbllx=0.7cm,bblly=5.9cm,bburx=20.5cm,bbury=19.5cm}\hss
  \epsfig{file=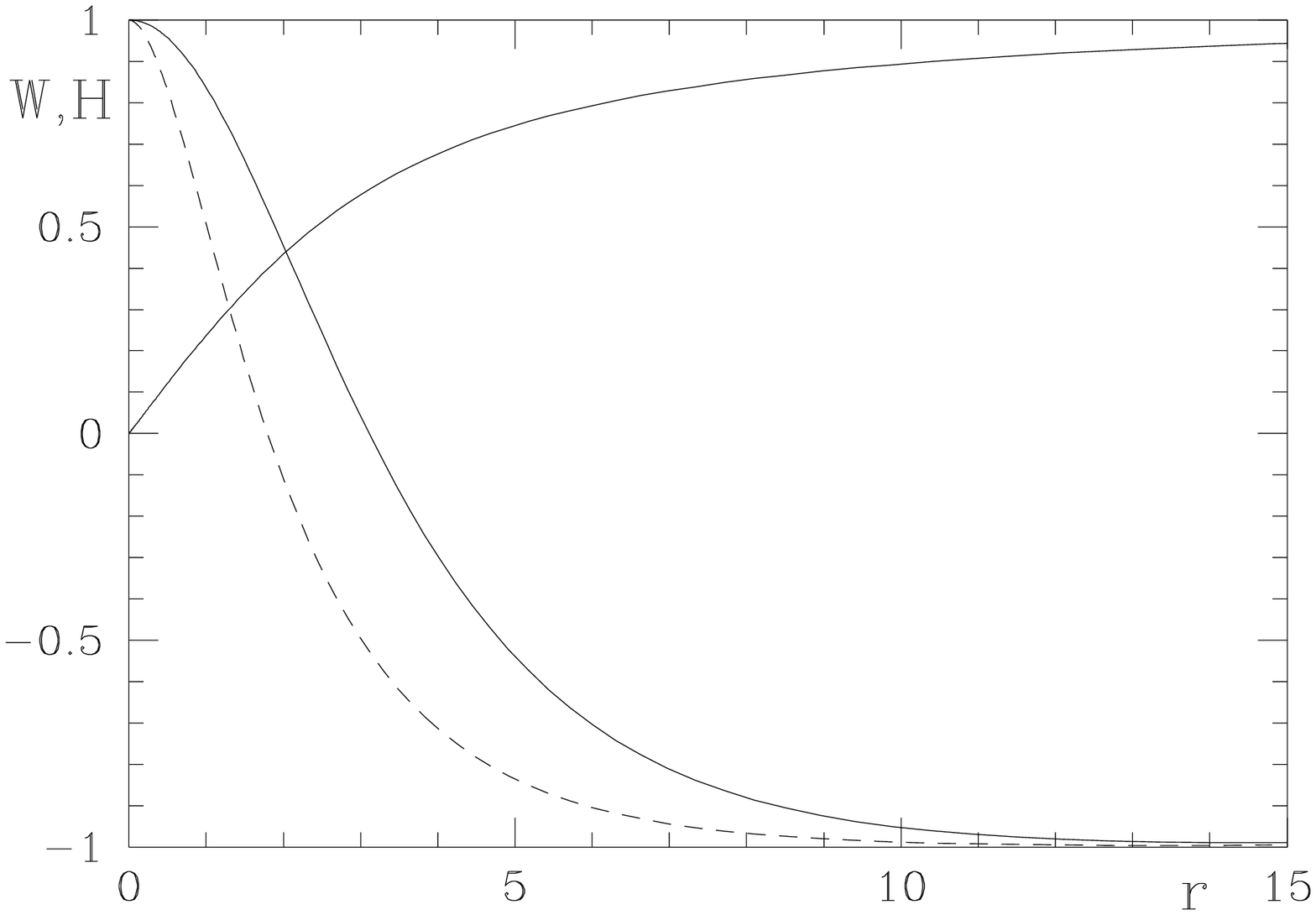,width=0.48\hsize,%
      bbllx=0.7cm,bblly=5.9cm,bburx=20.5cm,bbury=19.5cm}
  }
\caption[figflat]{\label{figflat}
a) PS-monopole, b) DHN-sphaleron, both for vanishing (solid) and infinite
(dashed) Higgs mass}   
\end{figure}

Before I come to the effects of gravity I would like to give 
a short reminder of ``particle like'' solutions of the YMH
system in flat space. For simplicity I restrict myself to the gauge group
{\sl SU(2)} from now on.
There are two different cases to be considered, leading to rather different
types of solutions.
The Higgs field can be either in a triplet or in a doublet representation.
In either case the action is
\begin{equation}\label{action}
S=-{1\over4\pi}\int d^4\;x\Bigl[{1\over4g^2}{\rm Tr}F^2+{1\over2}
|D\phi|^2+{\lambda\over8}(|\phi|^2-v^2)\Bigr]\;.
\end{equation}
It is important to notice that the expression for the action contains two
mass scales, the mass $M_W=gv$ of the YM field and the mass 
$M_H=\sqrt\lambda v$ of the Higgs field. From these we may form the
dimensionless ratio $\beta=M_H/M_W$.

The particle like (static, spherically symmetric) solutions 
in the case of a Higgs triplet are the 't~Hooft-Polyakov magnetic
monopoles. They are obtained with the ansatz
\begin{equation}\label{ansatz}
W^a_0=0\quad W^a_i=\epsilon_{iak}{x^k\over r}(W(r)-1)\quad 
      \phi^a={x^a\over r}H(r)\;.
\end{equation}
Inserting this ansatz in the action (\ref{action}) one gets
\begin{equation}\label{mono}
S=-\int dr\Bigl[{1\over g^2}(W'^2+{(1-W^2)^2\over2r^2})+{r^2\over 2}H'^2+
{\lambda r^2\over8}(H^2-v^2)^2+W^2H^2\Bigr]\;.
\end{equation}
In order to obtain finite total energy the Higgs field 
has to tend to its vacuum
value $v$ for $r\to\infty$, forcing in turn $W\to 0$ (l.h.s. of 
Fig.~\ref{figflat}).

For large values of $M_H$ and hence of $\beta$ the function $H(r)$ 
rises quickly to its asymptotic value $v$.
In the limit $\beta\to\infty$ the Higgs field may be replaced by $v$ for all 
$r>0$ and its only role is to give a mass to the YM field. The total energy
of the solution stays finite in this limit.
In fact, it only varies by a factor $\approx 1.8$ as $\beta$ varies from $0$
to $\infty$.

There is a second possibility to let $\beta$ go to infinity, holding $M_H$
fixed, but letting $M_W\to 0$ (and hence $W\equiv1$ at the relevant 
length scale $1/M_H$).
This way one obtains the ``global monopole'' playing the role of a texture
in cosmological considerations \cite{Vil}. 

Due to the topological character of the magnetic charge, related to the
asymptotic vacuum structure of configurations with finite energy,
the monopole is a stable solution.

The second possibility is a Higgs field in the doublet representation.
The relevant ansatz of the Higgs field is $\Phi^\alpha=H(r)\xi^\alpha$
with some constant spinor $\xi$.
Although this ansatz is not itself spherically symmetric it leads to a
consistent reduction. The corresponding reduced action is
\begin{equation}\label{sphal}
S=-\int dr\Bigl[{1\over g^2}(W'^2+{(1-W^2)^2\over2r^2})+{r^2\over 2}H'^2+
{\lambda r^2\over8}(H^2-v^2)^2+{1\over4}(W+1)^2H^2\Bigr]\;.
\end{equation}
The only essential difference of this action to the one for the triplet
is the form of the mass term. It destroys the symmetry $W\to -W$ and
enforces  $W$ to turn to $W=-1$ for $r\to\infty$ in order to have finite
total energy (Fig.~\ref{figflat}). 
This asymptotic behaviour implies that the solution has no
magnetic charge in contrast to the previous case with $W\to 0$.

In contrast to the stable monopole the sphaleron, i.e.\ the
solution minimizing the energy $H=-S$, is unstable. 
In order to understand this instability it is important to consider the
most general spherically symmetric ansatz for the YM field.
\begin{eqnarray}\label{sphericalYM}
W_t^a&=(0,0,A_0)\,,\qquad  W_\theta^a&=(W_1,W_2,0)\,\nonumber  \\
 W_r^a&=(0,0,A_1)\,, \qquad
 W_\varphi^a&=(-W_2\sin\theta,W_1\sin\theta,\cos\theta)\,.
\end{eqnarray}
The ansatz used above for the monopole and the sphaleron corresponds to a
consistent reduction putting $A_0=A_1=W_2=0$ and $W_1=W$.
The sphaleron turns out to be stable under variations staying within the
minimal reduction, but not if $\delta W_2\neq0$ and $\delta A_1\neq0$.
As was discussed by Manton \cite{Manton}
this instability is due to the non-trivial topology of the configuration
space of the spherically symmetric YM potential, again related to the 
asymptotic vacuum structure of configurations with finite energy.

\section{Gravitating Monopoles}\label{gmon}
\begin{figure}
\hbox to\hsize{
   \epsfig{file=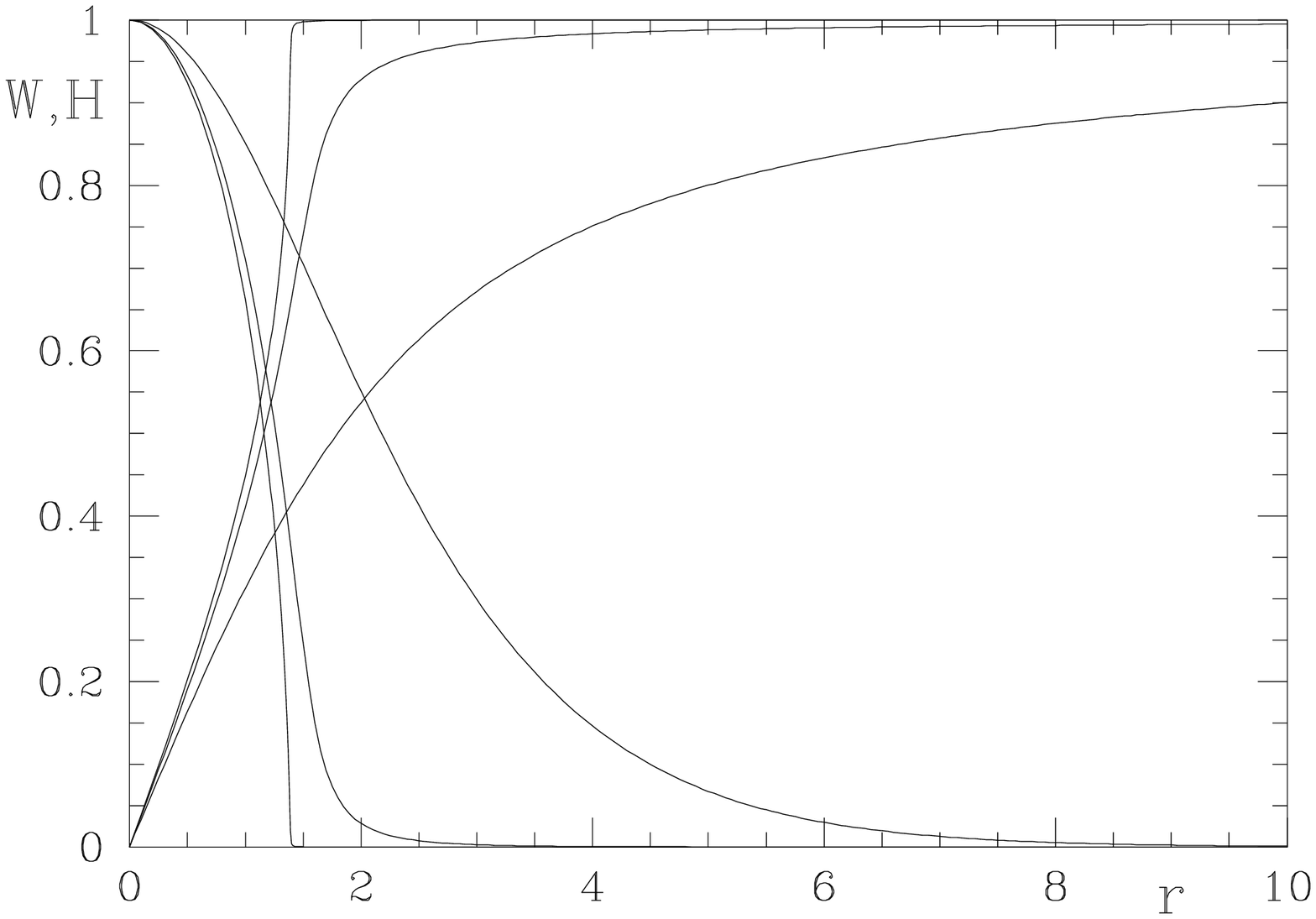,width=0.48\hsize,%
         bbllx=0.7cm,bblly=5.8cm,bburx=20.3cm,bbury=19.5cm}\hss 
   \epsfig{file=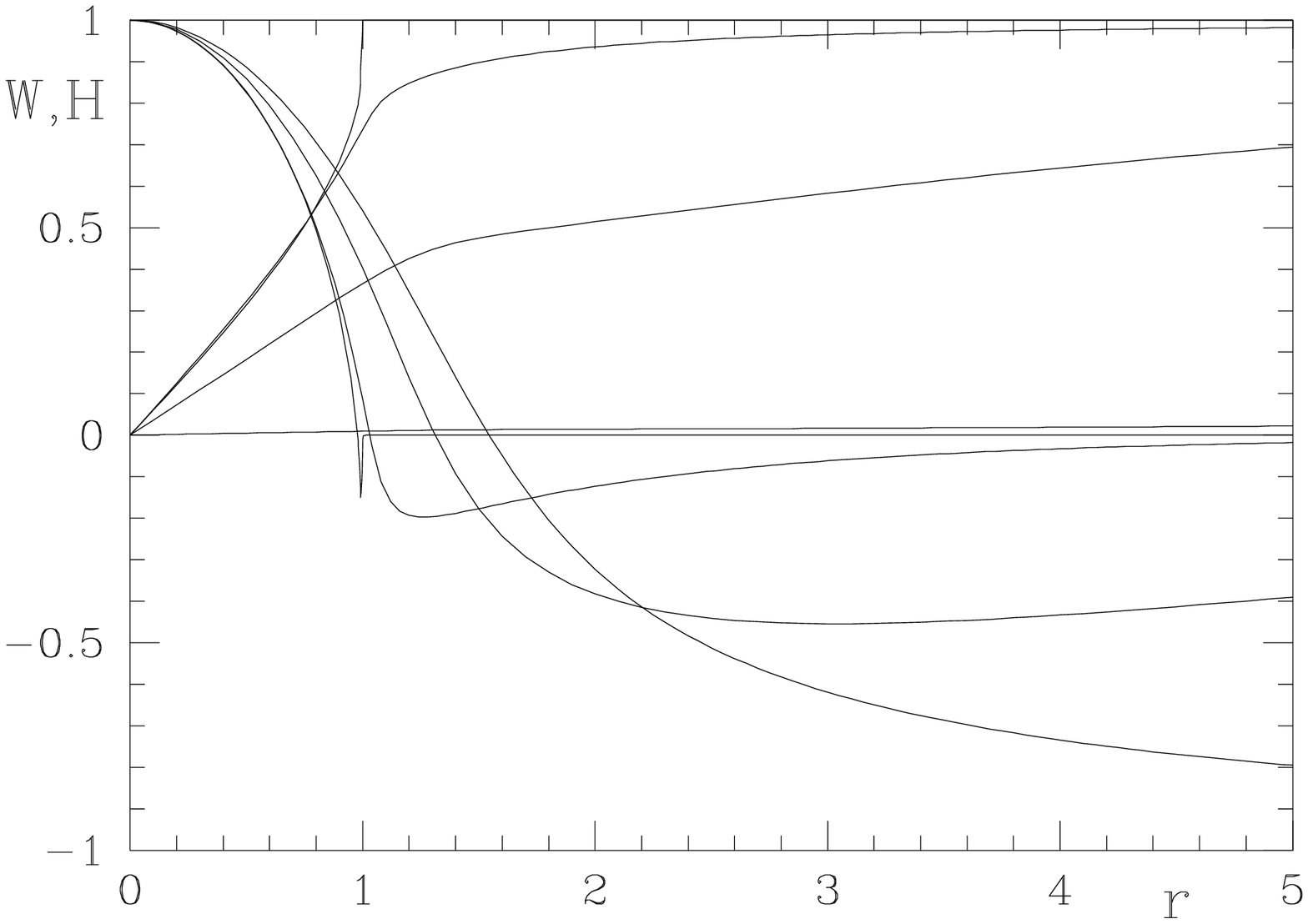,width=0.48\hsize,%
         bbllx=0.7cm,bblly=5.8cm,bburx=20.3cm,bbury=19.5cm}
  }
\caption[figmon]{\label{figmon}
W and H for ($\beta=0$) a) the gravitating monopole solutions for 
$\alpha=0.05,
\alpha_{\rm max}=1.403$ and $\alpha_c=1.386$; b) first radial excitation 
for $\alpha=0.01,0.2,0.5$ and $0.86$.}
\end{figure}

A spherically symmetric gravitational field is described by a 
space-time metric of the form
\begin{equation}\label{Metr}
ds^2=e^{2\nu}dt^2-e^{2\lambda}dR^2
  -r^2d\Omega^2\;,
\end{equation}
where $d\Omega^2=d\theta^2+\sin^2\theta d\varphi^2$ and $\nu,\lambda$ and $r^2$
are functions of the coordinates $t$ and $R$.

The quantity $r^2$ is proportional to the surface area of the
invariant 2-spheres and hence has a geometrical significance. 
Furthermore
for static space-times the function $e^\nu$ measures the invariant length
of the time-translation Killing vector, thus only 
the function $e^\lambda$ is gauge dependent, i.e., depends on
the choice of the radial coordinate $R$.

A simple gauge choice is $R=r$ (Schwarzschild coordinates), which is however 
well defined only as long as $dr/dR\neq0$.
Another convenient choice is obtained putting $e^\lambda=r$
(isotropic coordinates), leading to an autonomous form of the field
equations.

With gravity a new scale comes in through Newton's constant $G$, which allows us
to define the Planck mass $M_{\rm Pl}=1/\sqrt G$.
The existence of two different scales, the YM scale given by $M_W$ 
and the Planck scale $M_{\rm Pl}$ have a very important impact on the
structure of the solutions. In particular, studying the limiting case 
$M_{\rm Pl}>>M_W$ gives important insights for their interpretation, as we will
see in the following.   

Together with $M_W$ we can form the dimensionless ratio 
$\alpha=M_W\sqrt G/g=M_W/gM_{\rm Pl}$.
As just mentioned, a
special role is played by the limiting case $\alpha\to0$, which can however be
achieved in two different ways:
\begin{enumerate}
\item[i)] $G\to0$, $M_W$ fixed, in which the gravitational field decouples
(flat space);
\item[ii)] $v=M_W/g\to0$, $G$ fixed, in which the Higgs field becomes
trivial and can be ignored.
\end{enumerate}

The reduced EYMH action can be expressed as
\begin{equation}\label{gaction}
S=-\int dR e^{(\nu+\lambda)}
  \Bigl[
  {1\over2}\Bigl(1+e^{-2\lambda}((r')^2
   +\nu'(r^2)'\Bigr)- e^{-2\lambda}r^2V_1-V_2-V_3
\Bigr]\;,
\end{equation}
with
\begin{equation}
V_1={(W')^2\over r^2}+{1\over2}(H')^2\;,
\end{equation}

\begin{equation}
V_2={(1-W^2)^2\over2r^2}+
{\beta^2r^2\over8}(H^2-\alpha^2)^2
\end{equation}
and
\begin{equation}
V_3=W^2H^2 \quad {\rm resp.}\quad V_3={1\over4}(W+1)^2H^2
\end{equation}
for the triplet resp.\ doublet Higgs.
Through a suitable rescaling we have achieved that the action depends
only on
the dimensionless parameters $\alpha$ and $\beta$.
\begin{figure}
\hbox to\hsize{\hss
  \epsfig{file=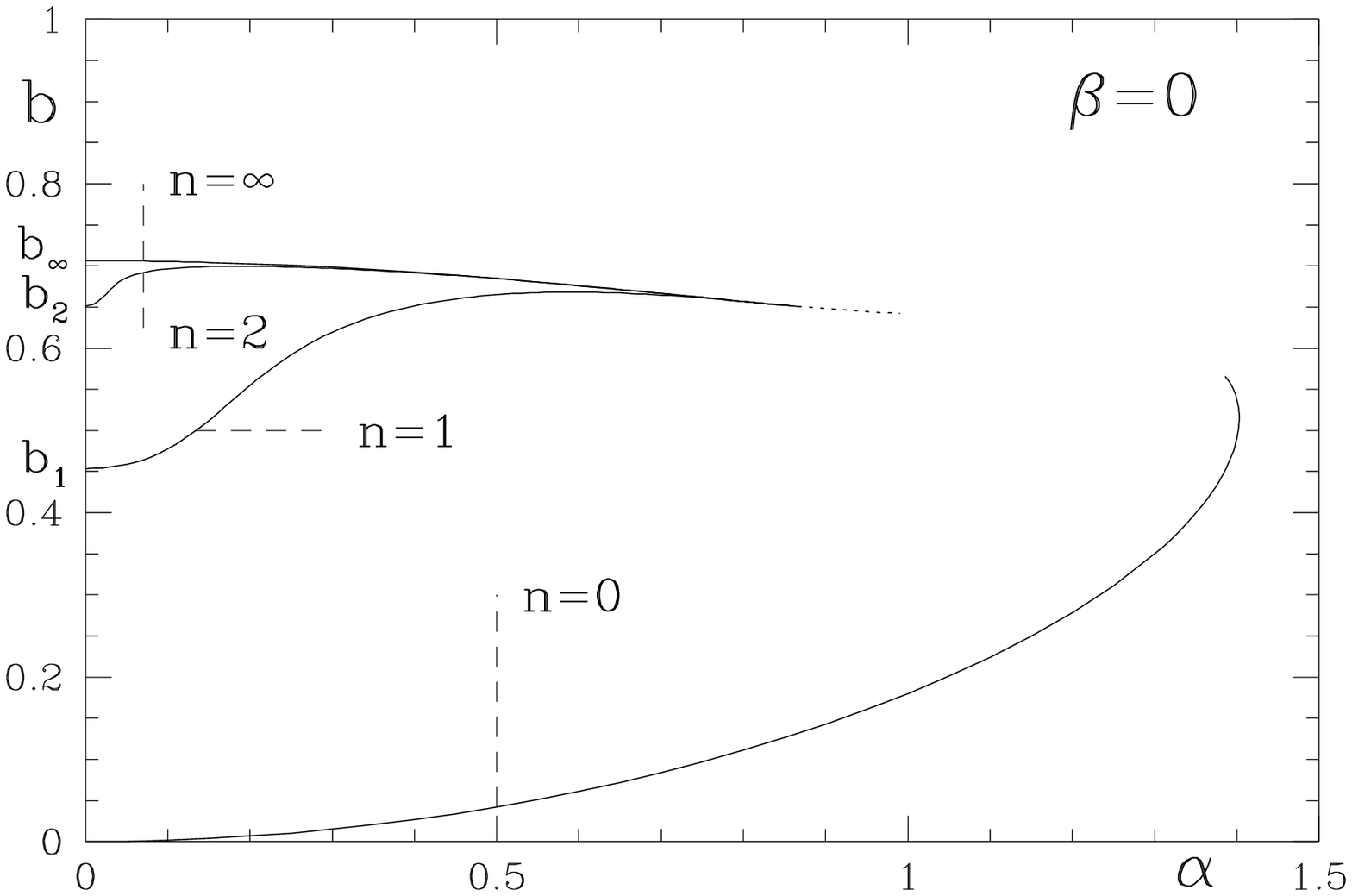,width=0.5\hsize,%
          bbllx=1.7cm,bblly=6.5cm,bburx=20cm,bbury=18.5cm}\hss
  \epsfig{file=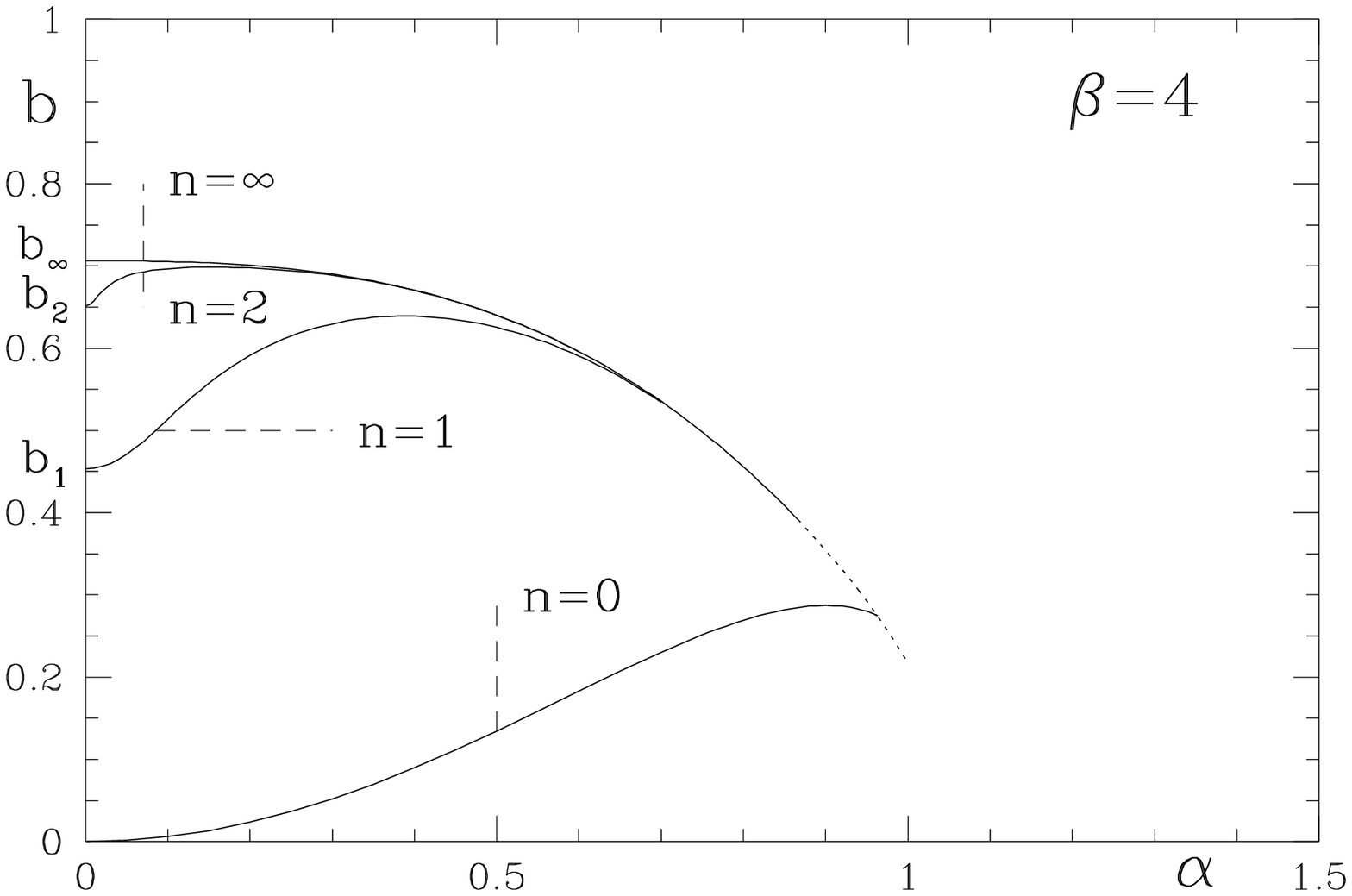,width=0.5\hsize,%
          bbllx=1.7cm,bblly=6.5cm,bburx=20cm,bbury=18.5cm}\hss
  }
\caption[figbalpha]{\label{figbalpha}
  The parameter $b$ vs.\ $\alpha$ of the fundamental monopole (lower branch),
 the first two excited ones, and the limiting solution ($n\to\infty$) for
  $\beta=0$ and $\beta=4$}
\end{figure}

Well-known exact solutions of the coupled Einstein-YM equations, obtained
from the action (\ref{gaction}),
are the Schwarzschild solution with trivial YM and Higgs fields
$W\equiv1,H\equiv v$
and the abelian (magnetically charged) Reissner-Nord\-str{\o}m (RN) solution with
$W\equiv0,H\equiv v$.
Both describe static black holes with a curvature singularity at the origin.
As was already mentioned, 
there exist no regular solitons of the 
Einstein-Maxwell theory.

Besides these trivial abelian
solutions there is a rich spectrum of non-abelian solutions found by
numerical integration of the field equations with suitable boundary
conditions \cite{BFM,Lee}.

To begin with there are the self-gravitating versions of the flat-space
non-abelian monopoles, which are recovered in the limit $\alpha\to0$
($G\to0$) (l.h.s.\ of Fig.~\ref{figmon}). 
These are globally regular solutions
with a regular center of symmetry (origin) and finite mass.
They exist only
up to some maximal value $\alpha_{\rm max}$ (depending on $\beta$) 
of the mass ratio 
$\alpha=M_W/gM_{\rm Pl}$.
Such a maximal value of $\alpha$ is to be anticipated. We expect the monopole
to become gravitationally unstable, when its size 
$R_{\rm mon}\approx 1/M_W$ becomes comparable to its Schwarzschild radius
$R_{SS}=GM_{\rm mon}\approx GM_W/g^2$, i.e.\ for
$GM_W^2/g^2=\alpha^2\approx 1$.
As $\alpha$ increases 
the solutions develop a typical 
limiting behaviour indicating this instability,
which may be characterized as ``gravitational confinement''
of the monopole. The spatial hyper-surface $t={\rm const.}$ develops an
infinite throat separating an interior region with a smooth origin and 
non-trivial YM field from an exterior extremal RN solution with $W\equiv 0$.
This throat is characterized by a finite limiting value $r_l=\alpha$
of the metric function $r$. 
Geometrically this means that neighbouring radial light-rays become
non-divergent. All this is much like the $t={\rm const.}$ surfaces of the
extremal RN solution, with the only difference, that the interior part of
the throat is not the analytic continuation of the exterior one. In fact the
combination of the 
metric functions $\nu+\lambda$ blows up along the 
throat coming from the interior, whereas $\nu+\lambda\equiv0$ for the RN
solution. 

Amazingly, for small values ($\lsim 0.7$)
of the parameter $\beta=M_H/M_W$ this kind of 
singular limiting behaviour does not occur at the maximal value of $\alpha$
but at some critical value 
$\alpha_c<\alpha_{\rm max}$. Running along the 1-paramter family of
solutions starting at $\alpha=0$ and increasing $\alpha$ one runs through
a maximum of $\alpha$ before the slightly smaller 
critical value $\alpha_c$ is reached, i.e.\
there exist two monopole solutions to the same value of $\alpha$ in the
interval $\alpha_c<\alpha<\alpha_{\rm max}$. This double-valuedness can be
avoided using a different parameter for this family of solutions. 
For solutions with a regular origin the YM potential $W$ has the behaviour
$W=1-br^2+O(r^4)$ for $r\to 0$. Given $\alpha$ the parameter $b(\alpha)$ is
fixed by the requirement of asymptotic flatness. It turns out that the
parametrization with $b$ instead of $\alpha$ is one-to-one
(compare Fig.~\ref{figbalpha}).
\begin{figure}
\hbox to\hsize{
   \epsfig{file=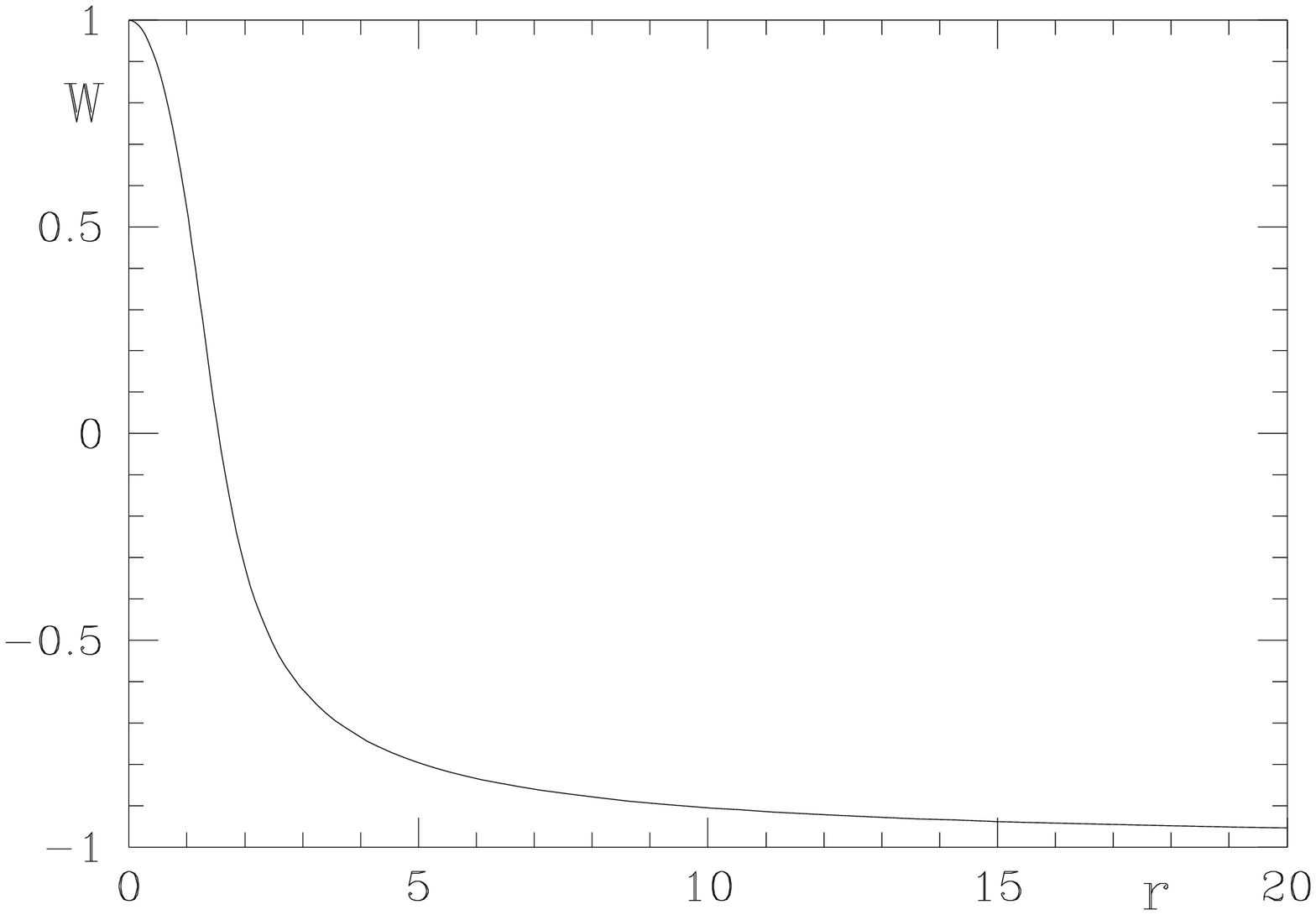,width=0.48\hsize,%
         bbllx=0.7cm,bblly=5.9cm,bburx=20.5cm,bbury=19.5cm}\hss 
   \epsfig{file=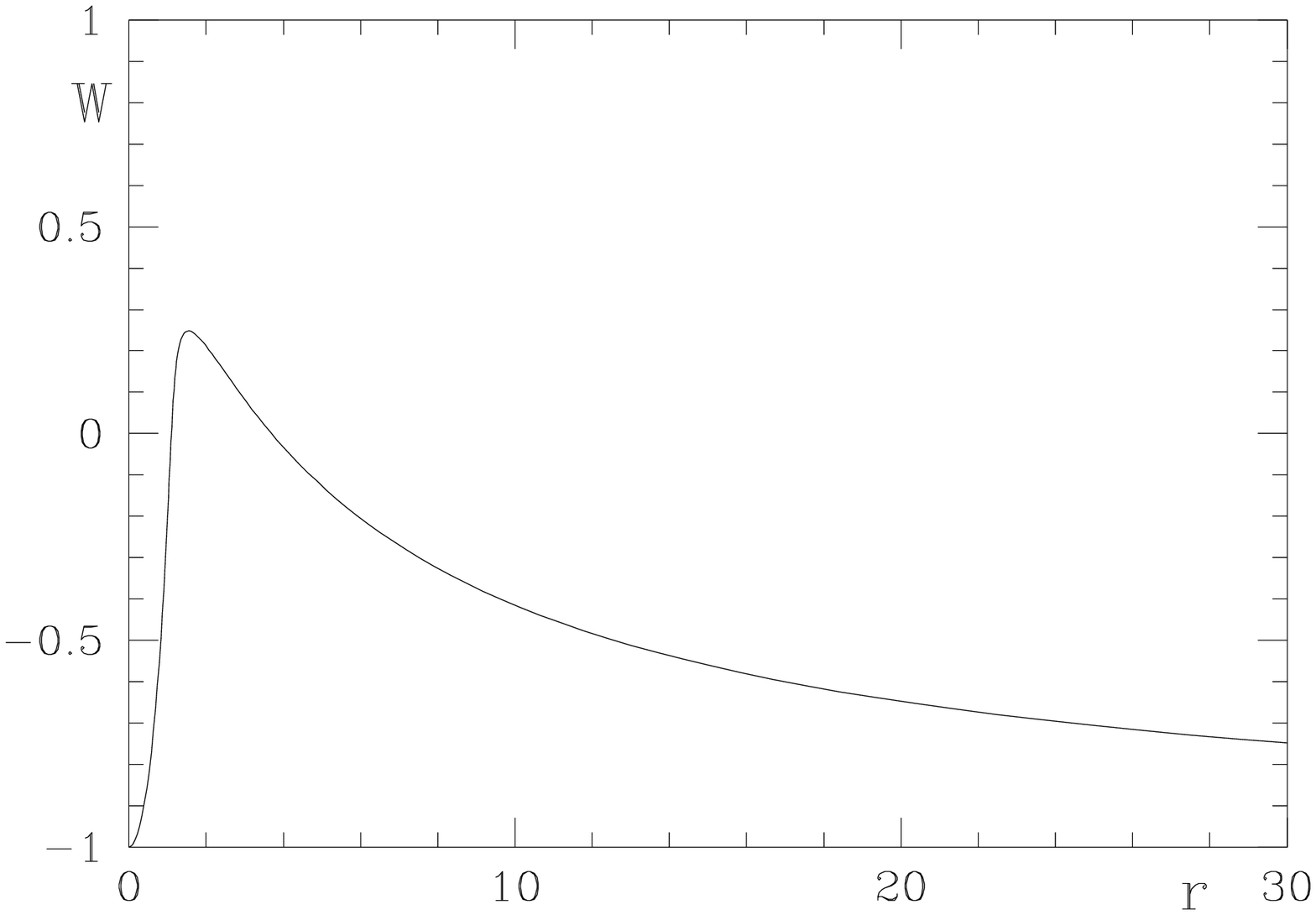,width=0.48\hsize,%
         bbllx=0.7cm,bblly=5.9cm,bburx=20.5cm,bbury=19.5cm}
  }
\caption[figsph]{\label{figbart}
W for the first two Bartnik-McKinnon solutions 
}
\end{figure}

This figure contains also the first two members of a 
sequence of excited families of monopoles 
showing a different behaviour of $b(\alpha)$ for $\alpha\to 0$. The
corresponding values of $b$ tend to finite values $b_n$ ($n=1,2,...$)
as $\alpha$ tends to 0 ($v\to0$) related to the solutions found by Bartnik and
McKinnon (BM).
The latter are globally regular solutions of the EYM equations without a
Higgs field. They are labelled by the number $n$ of zeros of the YM
potential $W$. Their mass $M$ 
is of order one in units of $M_{\rm Pl}/g$ --- the only mass
scale in this case --- tending rapidly to $M=1$ for growing $n$. 
The parameters $b_n$ converge to $b_\infty\approx 0.7064$.
In contrast to
the monopole solutions they carry zero magnetic charge, related to their
different asymptotic behaviour for $r\to\infty$, where $W\to\pm1$.
(Compare Fig.~\ref{figbart}). The shape of the $n=1$ solution reminds very much of the
flat space sphaleron. In fact, it is also unstable and may be understood as
a gravitationally bound sphaleron, the gravitational field replacing the
Higgs field of the flat sphaleron \cite{Wald}.
That this interpretation makes good sense is underlined by the fact that
similar solutions are obtained with a scalar dilaton replacing the
gravitational field \cite{LM}.

For finite, but small $\alpha$ the excited monopoles consist essentially of
a very small (Planck size) BM-solution sitting inside a large (size
$1/M_W$) flat monopole (r.h.s. of Fig.~\ref{figmon}). Let me recall that for
all the monopole solutions $W\to0$ for $r\to\infty$, even though this is not
clearly visible for all the curves of the plot.  

All these families of excited monopole solutions
have a common value of $\alpha_c=\sqrt3/2$, which is also the
maximal one in this case. For $\alpha\to\alpha_c$ we observe the same
limiting behaviour (infinite throat) as for the fundamental monopole. Thus
it seems that the latter describes a rather universal phenomenon indicating
gravitational instability of static equilibrium configurations
and hence is to be expected to occur also for other
gravitating matter systems.
\begin{figure}
\hbox to\hsize{
\epsfig{file=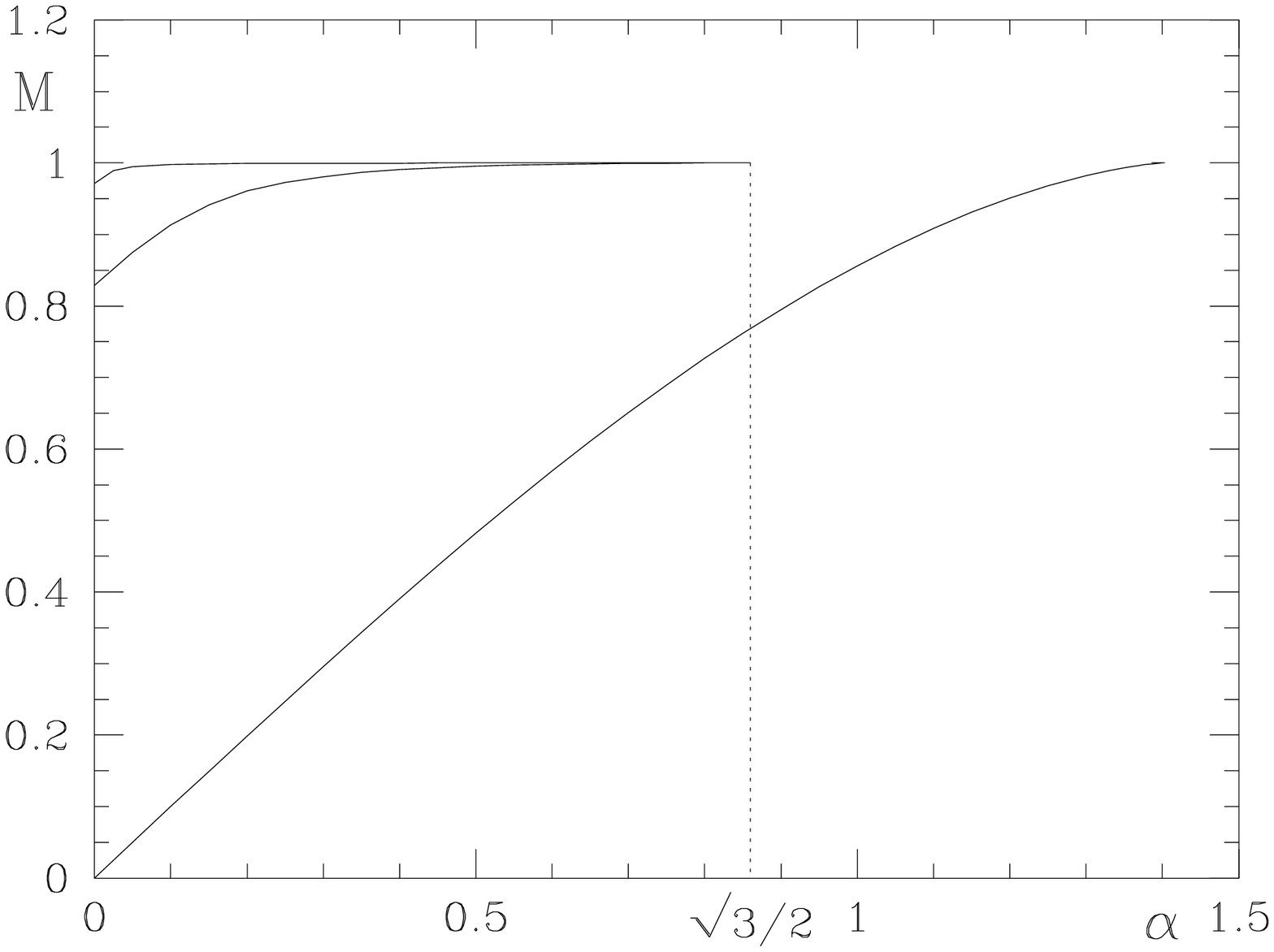,width=0.48\hsize,%
    bbllx=1.2cm,bblly=5.8cm,bburx=20.5cm,bbury=20cm}\hss
  \epsfig{file=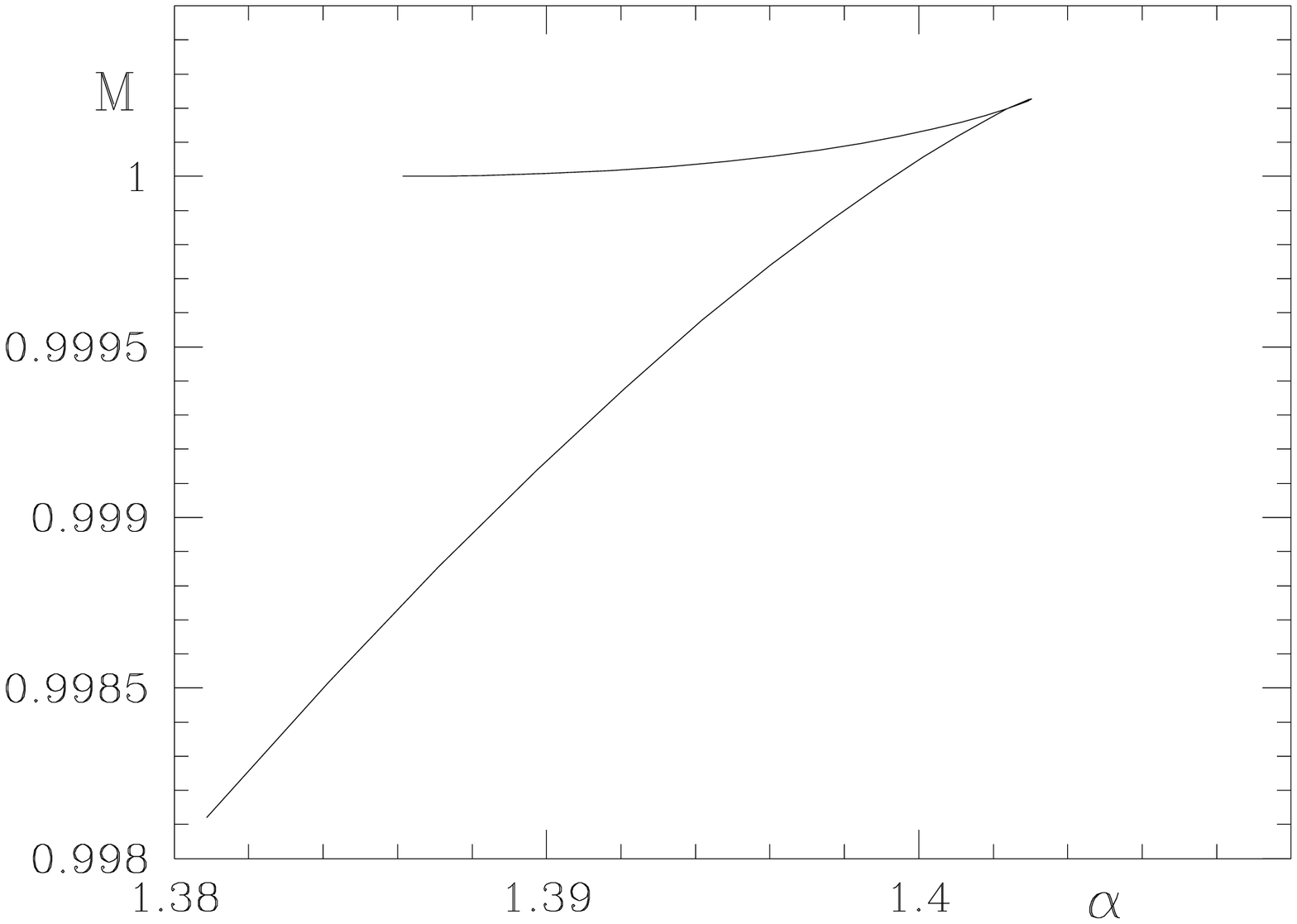,width=0.48\hsize,%
    bbllx=0.7cm,bblly=5.8cm,bburx=20.6cm,bbury=20cm}
  }
\caption[figmass]{\label{figmass}
a)Masses (in units of $M_{rm Pl}$/g) of 
fundamental monopole solutions and first
and second radial excitations versus $\alpha$ (for $\beta=0$);
b) the critical region for the fundamental
        solutions in detail.}
\end{figure}

The l.h.s.\ of
Fig.\ref{figmass} shows the masses of the 
various monopole solutions as a function of
$\alpha$ ($\beta=0$). We find that for $\alpha=\alpha_c$ the mass of the solutions
becomes $M=M_c\equiv M_{\rm Pl}/g$, the mass of the extremal RN solution.
This is
easily understood from the merging of the exterior throat part of the
monopoles with the latter solution as $\alpha\to\alpha_c$.
As indicated in the r.h.s.\ of  Fig~\ref{figmass}
the mass at $\alpha_{\rm max}$ is slightly bigger than $M_c$. 
While $\beta$ increases $\alpha_c$ decreases to the limiting value 
$\alpha_c=\sqrt{2}/2$ for $\beta\to\infty$.
Similar to their flat counterparts solutions for $\beta=\infty$ 
(i.e.\ $H\equiv v$) may be considered as cosmological textures.

\section{Non-abelian Black Holes}\label{bh}
\begin{figure}
\hbox to\hsize{
   \epsfig{file=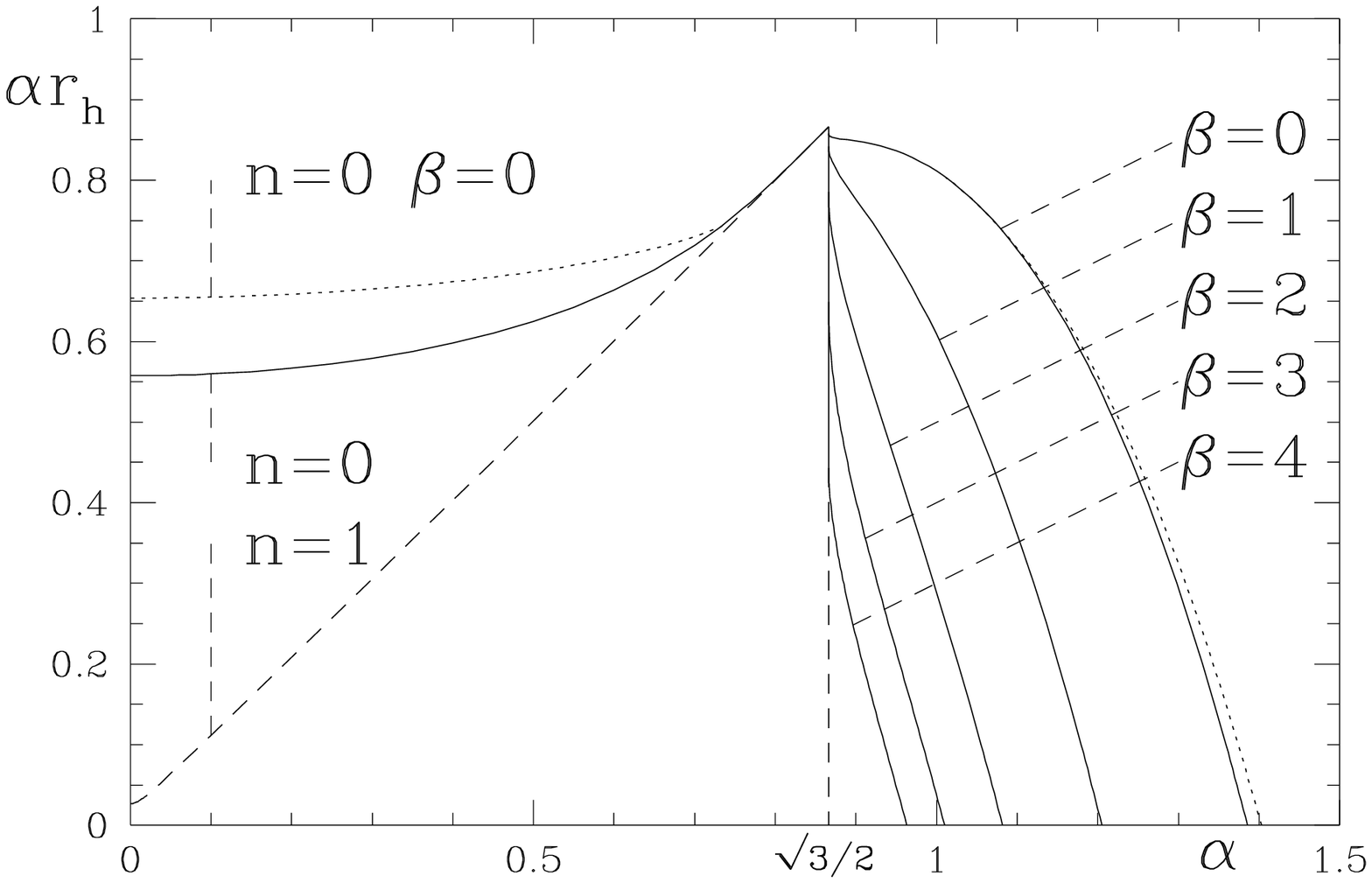,width=0.48\hsize,%
   bbllx=1.1cm,bblly=6.3cm,bburx=20cm,bbury=18.5}\hss
   \epsfig{file=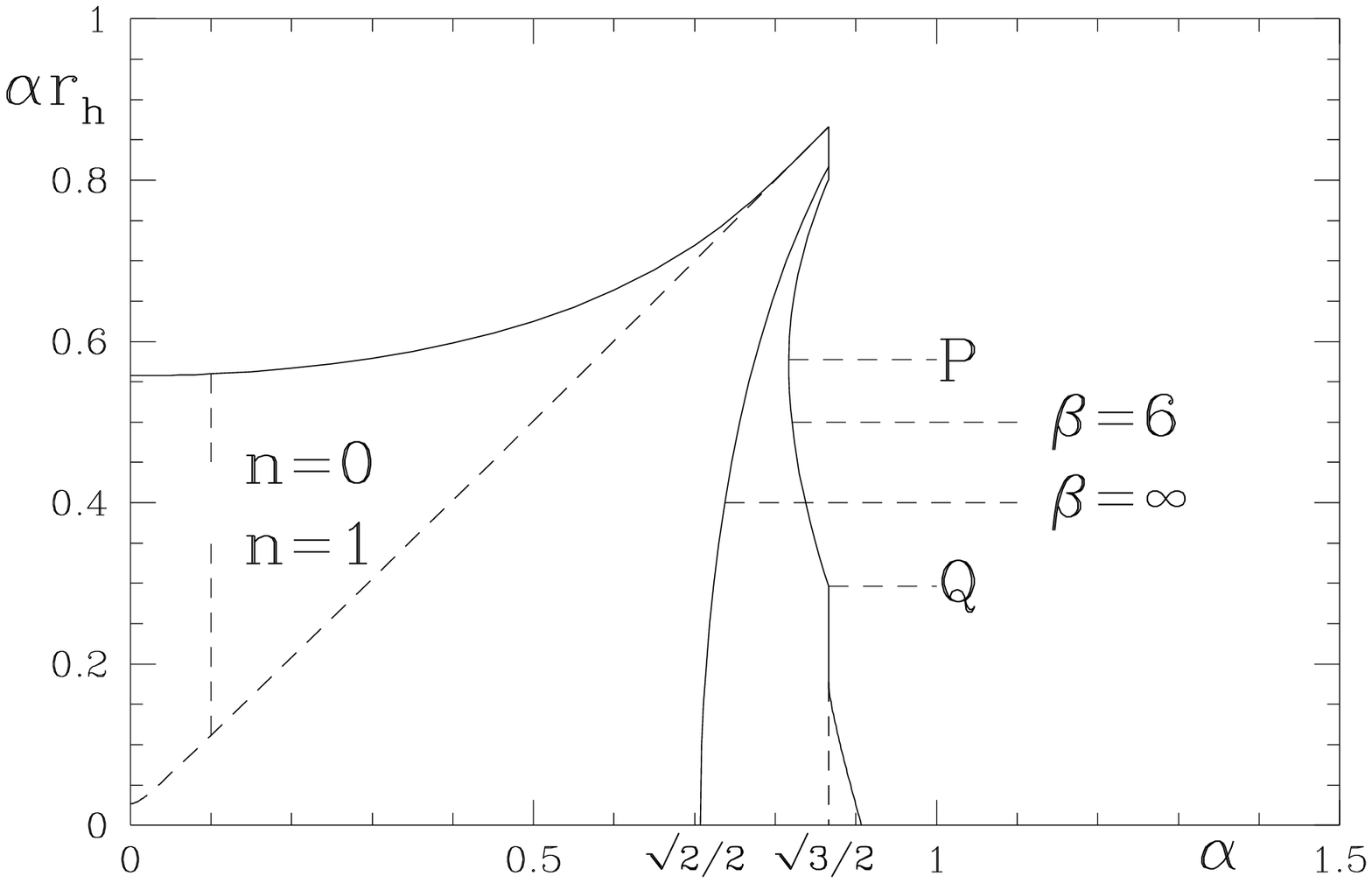,width=0.48\hsize,%
   bbllx=1.1cm,bblly=6.3cm,bburx=20cm,bbury=18.5cm}
  }
\caption[figdomain]{\label{figdomain}
  Domains of existence for non-abelian black holes:
  a) for $\beta=0$, $1$, $2$, $3$, and $4$;
  b) for $\beta=6$ and $\infty$}
\end{figure}

Apart from the solutions with a regular origin there are non-abelian,
``coloured'' black
holes, parametrized by their radius $r_h$ (in geometrical units, i.e.
the value of $r$ at the event horizon) in addition to $\alpha$ and $\beta$
\cite{BFM,Aichel}.
For $r_h<< 1/M_W$ these non-abelian black holes may be interpreted as a tiny
Schwarzschild black hole sitting inside a monopole.
On the other hand, when $r_h$ becomes bigger than $\approx 1/M_W$ this type
of solution disappears and only the abelian RN black holes exist.
For $r_h\to 0$ the matter fields tend uniformly to those of the globally regular
solutions, whereas for the metrical functions this limit is clearly more
delicate.

Detailed numerical analysis reveals that non-abelian black holes 
exist only in a limited domain of the $\alpha$-$r_h$-plane, whose shape
undergoes some characteristic changes as $\beta$ varies from 0 to $\infty$.
Fig.~\ref{figdomain} shows these domains. 
Observe that we use $\alpha r_h$ instead of $r_h$
as the abscissa - equivalent to expressing $r_h$ in units of $1/M_W$ - in
order to obtain domains remaining bounded for $\alpha\to0$.
In the following I shall discuss in some more detail the structure of these 
 ``Phase Diagrams'' and the phenomena happening at their boundaries.
Let me start with the case $\beta=0$.

\begin{figure}
\begin{center}
   \epsfig{file=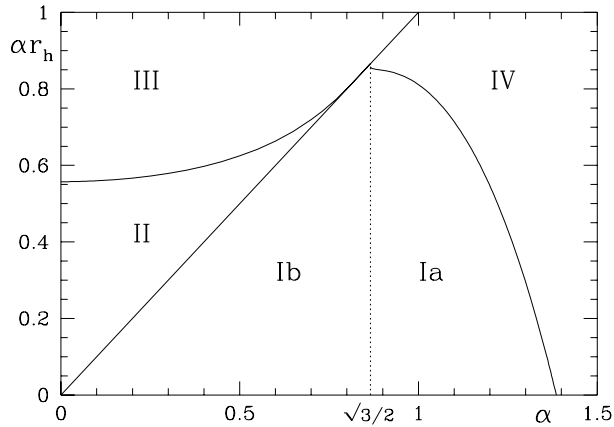,width=0.5\hsize,%
   bbllx=1.1cm,bblly=5.9cm,bburx=20.5cm,bbury=19.5cm}
\end{center}
\caption[figphase]{\label{figphase}
  Domains of existence for abelian and  non-abelian black holes}
\end{figure}

It is appropriate to subdivide the relevant sector $\alpha\geq0$, $r_h\geq0$
into the four subregions I-IV (compare Fig.~\ref{figphase}).

In regions I and II we find coloured black holes.
Above the diagonal, i.e.\ in regions II and III we have the abelian RN
solutions, the extremal RN black holes sitting on the diagonal.
Below the diagonal the RN solution has a naked singularity and does not
represent a black hole.
No black holes neither abelian nor non-abelian could be found in region IV.
Region I may be subdivided in region ${\rm I_a}$, where 
only the b.h.\ version of
the fundamental monopole resides and region ${\rm I_b}$, where 
in addition their radial
excitations are found. Thus region ${\rm I_a}$ contains essentially one
b.h.\ solution for given values of $\alpha$ and $r_h$ - apart from a small
interval $\alpha_c(r_h)<\alpha<\alpha_{\rm max}(r_h)$, where two solutions
exist) - whereas in region ${\rm I_b}$ countably many solutions exist
(for given $\alpha$ and $r_h$).

In region II abelian and non-abelian black holes coexist.
This establishes an obvious violation of the so-called ``No-Hair Conjecture''.
According to the latter black holes should (apart from mass and
angular-momentum) be uniquely determined through their ``gauge charges'' -
their magnetic charge in the present case. 
However, abelian and non-abelian black holes carry the same magnetic charge
and can also be made degenerate in mass resp.\ the value of $r_h$.
Because in general only one type of the ``degenerate'' black holes is stable
(compare below), a weakened form of the ``No-Hair Conjecture'',
including the requirement of stability, could be maintained. 

As $\beta$ increases from $0$ to $\beta=4$
the structure of the ``Phase Diagram'' remains essentially the same, the right
boundary curve moving in to the left.
However, for $\beta>4$ this boundary curve 
develops a second, concave branch (compare Fig.~\ref{figdomain}) 
determined by another mechanism, the formation of a degenerate (inner
or outer) horizon. 

The boundary curve above the diagonal is essentially
characterized by the bifurcation of
the non-abelian with the abelian RN solution. 
For a given value of $\alpha$ this 
happens at some value $r_{\rm h,c}(\alpha)$.
Approaching this value
from below the value $W_h$ of $W$ at the horizon tends to zero.

\begin{figure}
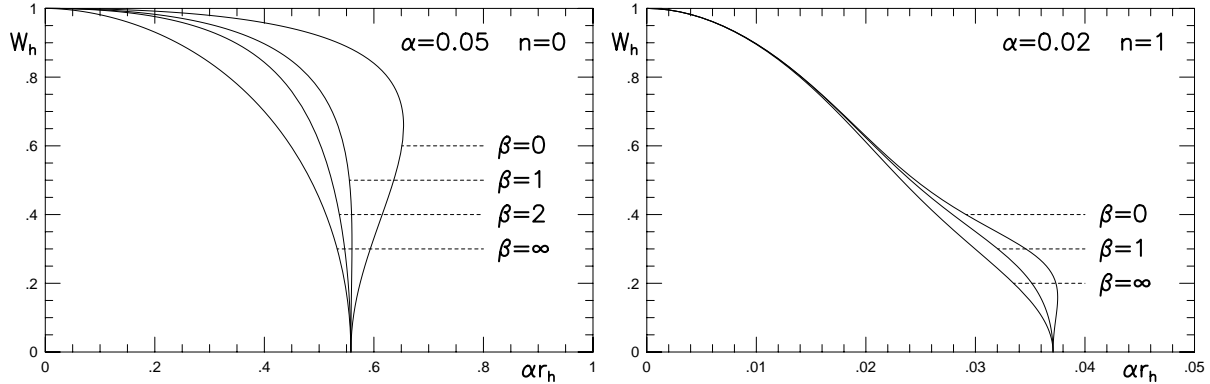

\hbox to\hsize{\hss
   \epsfig{file=rhwh0.eps,width=0.5\hsize,%
   bbllx=56bp,bblly=72bp,bburx=782bp,bbury=556bp}\hss
   \epsfig{file=rhwh1.eps,width=0.5\hsize,%
   bbllx=56bp,bblly=72bp,bburx=782bp,bbury=556bp}\hss 
  }
\caption[figrhwh]{\label{figrhwh}
  Initial data $r_h$ and $W_h$ for the fundamental and first excited
  black hole solution for $\alpha=0.05$, resp.\ $\alpha=0.02$ and
  various values of $\beta$}
\end{figure}

But again there is a slight complication for small values of 
$\beta$ ($\beta\lsim 1.1$);
similar to the existence of $\alpha_{\rm max}>\alpha_c$ there is a 
$r_{\rm h,max}>r_{\rm h,c}$-phenomenon (compare Fig.~\ref{figrhwh}).

\section{Stability of monopoles and coloured black holes}\label{stabmon}

I shall discuss here only stability against infinitesimal, spherically
symmetric perturbations.
In view of the time-independence of the solutions
this amounts to analyzing the spectrum of perturbations with harmonic
time-dependence obeying suitable boundary conditions.
Imaginary frequencies correspond to unstable modes of the solution.

As to be expected all the excited regular monopoles turn out to be unstable. 
The branch of gravitating monopoles connected to the flat space
solution is stable up to $\alpha_{\rm max}$, whereas the corresponding upper
branch - existing for $\alpha\epsilon[\alpha_c,\alpha_{\rm max}]$ is
unstable \cite{Helia}. This change of stability at the bifurcation point at 
$\alpha_{\rm max}$ of the massfunction is in agreement with general results on 
1-parameter families of solutions and well-known from stellar models
\cite{Wheel}.

Analogous results hold for the non-abelian, magnetically charged black holes.
It is, however, interesting to observe that the abelian RN black hole is
unstable in the framework of the non-abelian theory for 
$\alpha$ smaller than some value $\alpha(r_h)<\sqrt3/2$ \cite{Wein,BFM}.
In particular, the extremal RN solution is unstable for $\alpha<\sqrt3/2$
and stable above this value.
At the limiting value $\alpha=\sqrt3/2$ the extremal RN solution bifurcates
with infinitely many non-abelian solutions and in fact develops infinitely
many unstable modes.

\section{Gravitating Sphalerons and Sphaleron Black Holes}\label{gsphal}

In complete analogy to gravitating monopoles one may also consider
self-gravitating sphalerons, i.e. gravitating versions of the flat-space
sphaleron \cite{Greene}. 
Although many of the phenomena discussed above for monopoles 
repeat itself in this case, there are some important differences as far as
the domains of existence for sphaleron black holes are concerned. Also the
stability properties are clearly different.
\begin{figure}
\hbox to\hsize{
   \epsfig{file=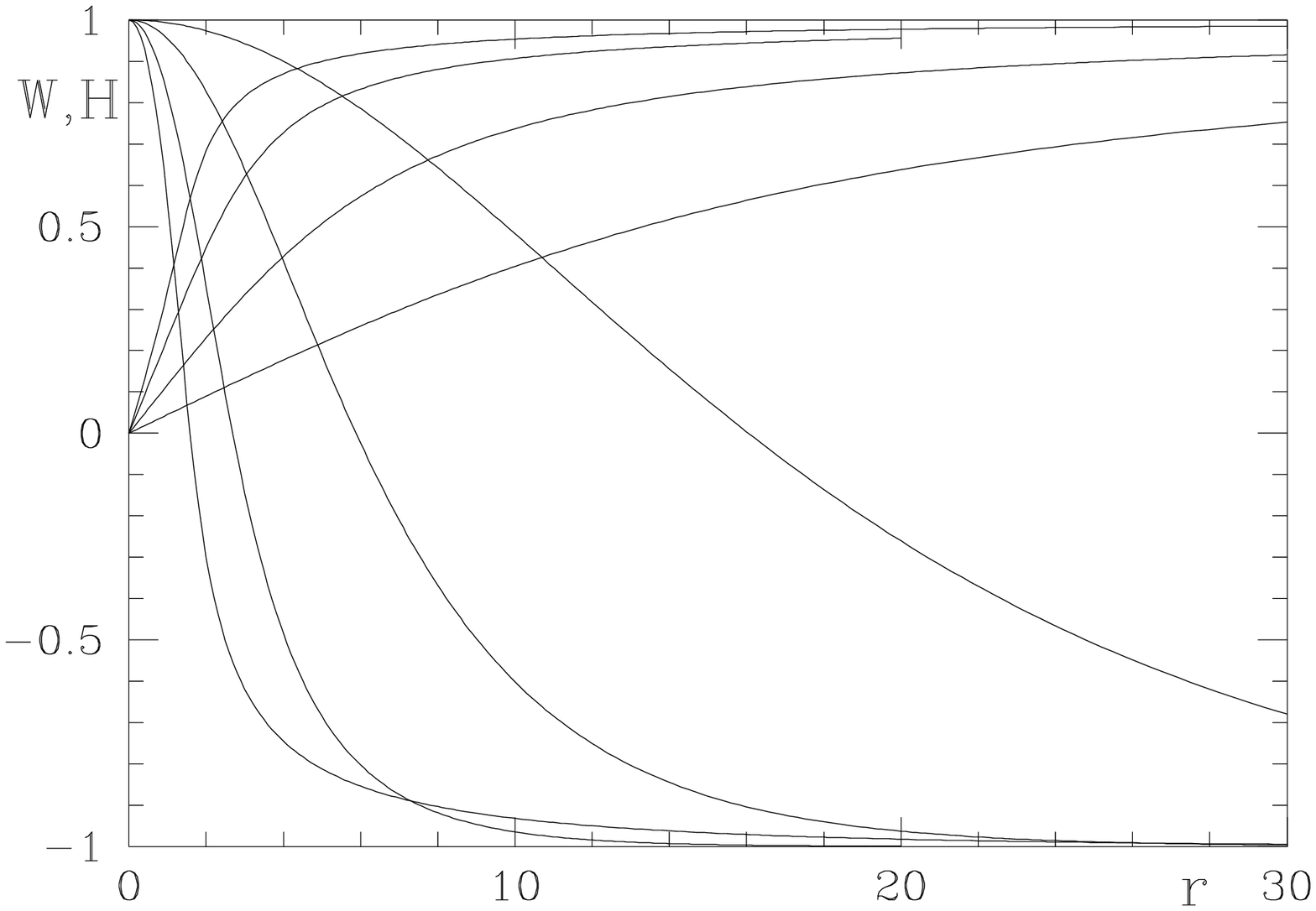,width=0.48\hsize,%
         bbllx=0.7cm,bblly=5.9cm,bburx=20.5cm,bbury=19.5cm}\hss 
   \epsfig{file=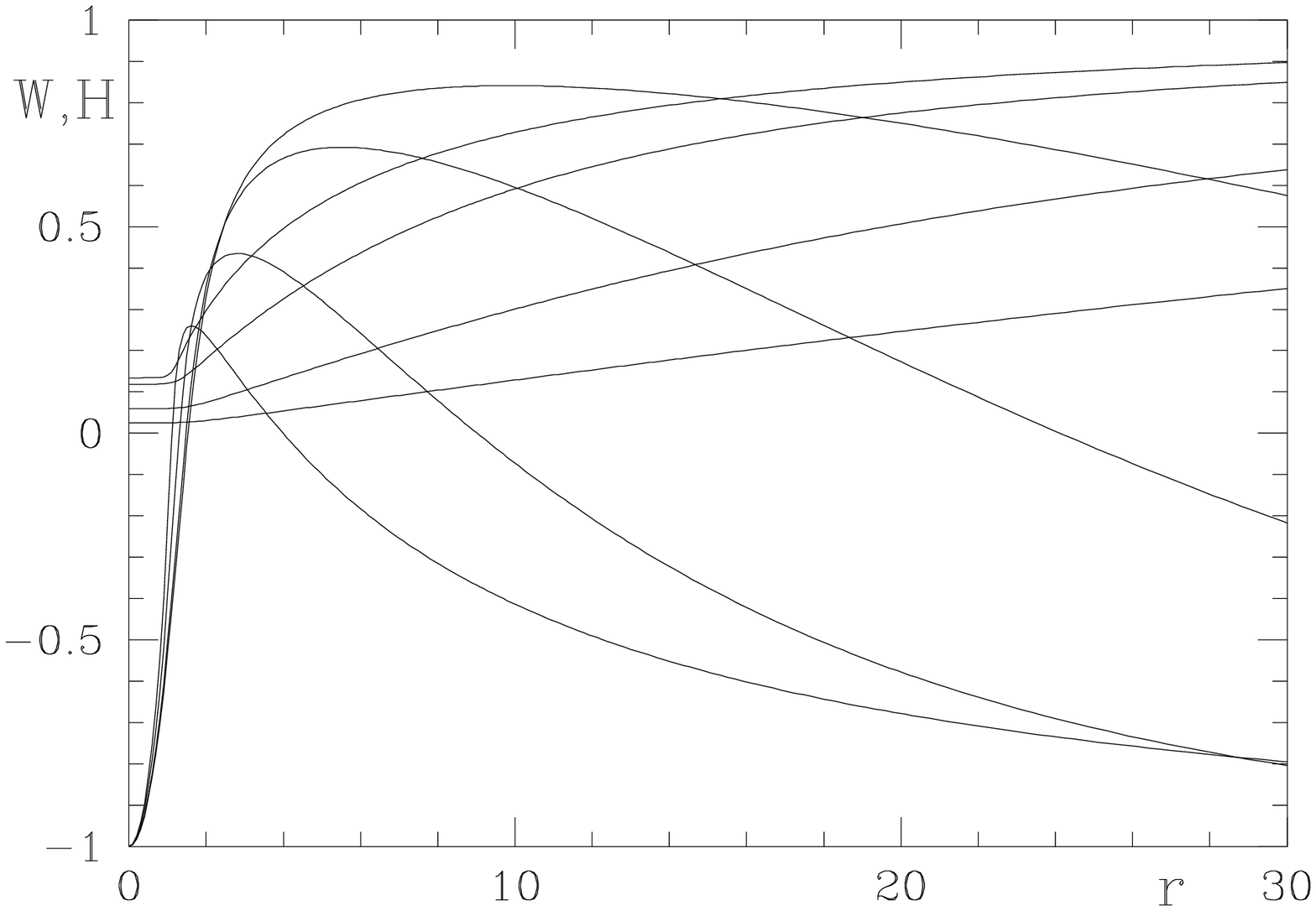,width=0.48\hsize,%
         bbllx=0.7cm,bblly=5.9cm,bburx=20.5cm,bbury=19.5cm}
  }
\caption[figsph]{\label{figsph}
W and H for the gravitating sphaleron solutions ($\beta=0$)
a) fundamental solution, $\alpha=0.2, 0.5, \alpha_{\rm max}=0.7478, 0.2$; 
b) first radial excitation, $\alpha=0.05, 0.1, \alpha_{\rm max}=0.144,
0.05$.}
\end{figure}
Again there is a maximal value of the parameter $\alpha$ for which static
gravitating sphalerons exist. However, it is a rather different mechanism
that is responsible for its existence.
For very small values of $\alpha$ the coupling to 
gravity yields only a small
perturbation of the flat-space sphaleron, whose size is $\approx 1/M_W$.
As $\alpha$ increases the gravitational attraction leads to shrinking of the
sphaleron, until it eventually becomes of size $\approx 1/M_{\rm Pl}$
(l.h.s.\ of Fig.~\ref{figsph}).

\begin{figure}
\hbox to\hsize{
  \epsfig{file=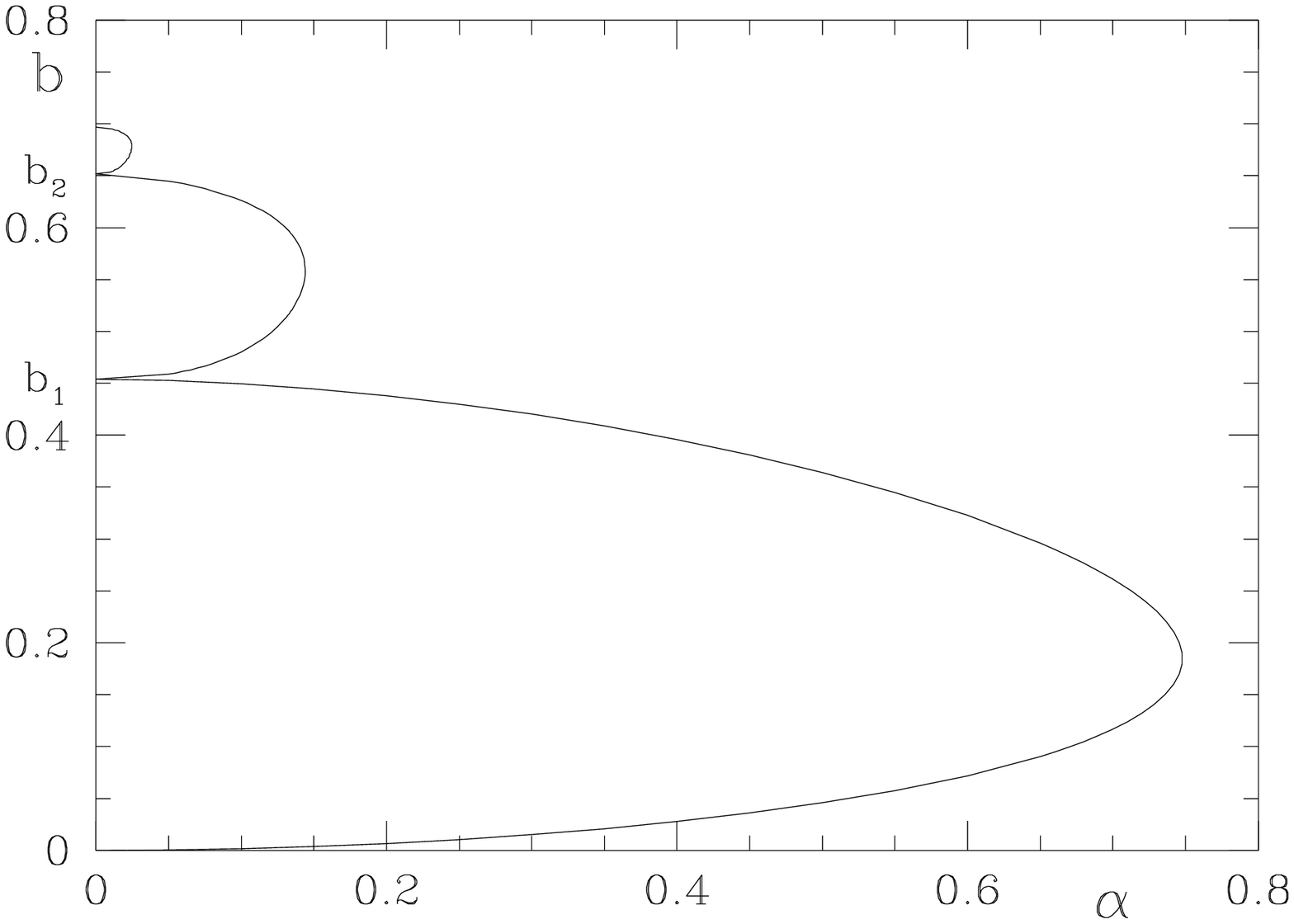,width=0.48\hsize,%
          bbllx=1.1cm,bblly=5.8cm,bburx=20.5cm,bbury=19.5cm}\hss
  \epsfig{file=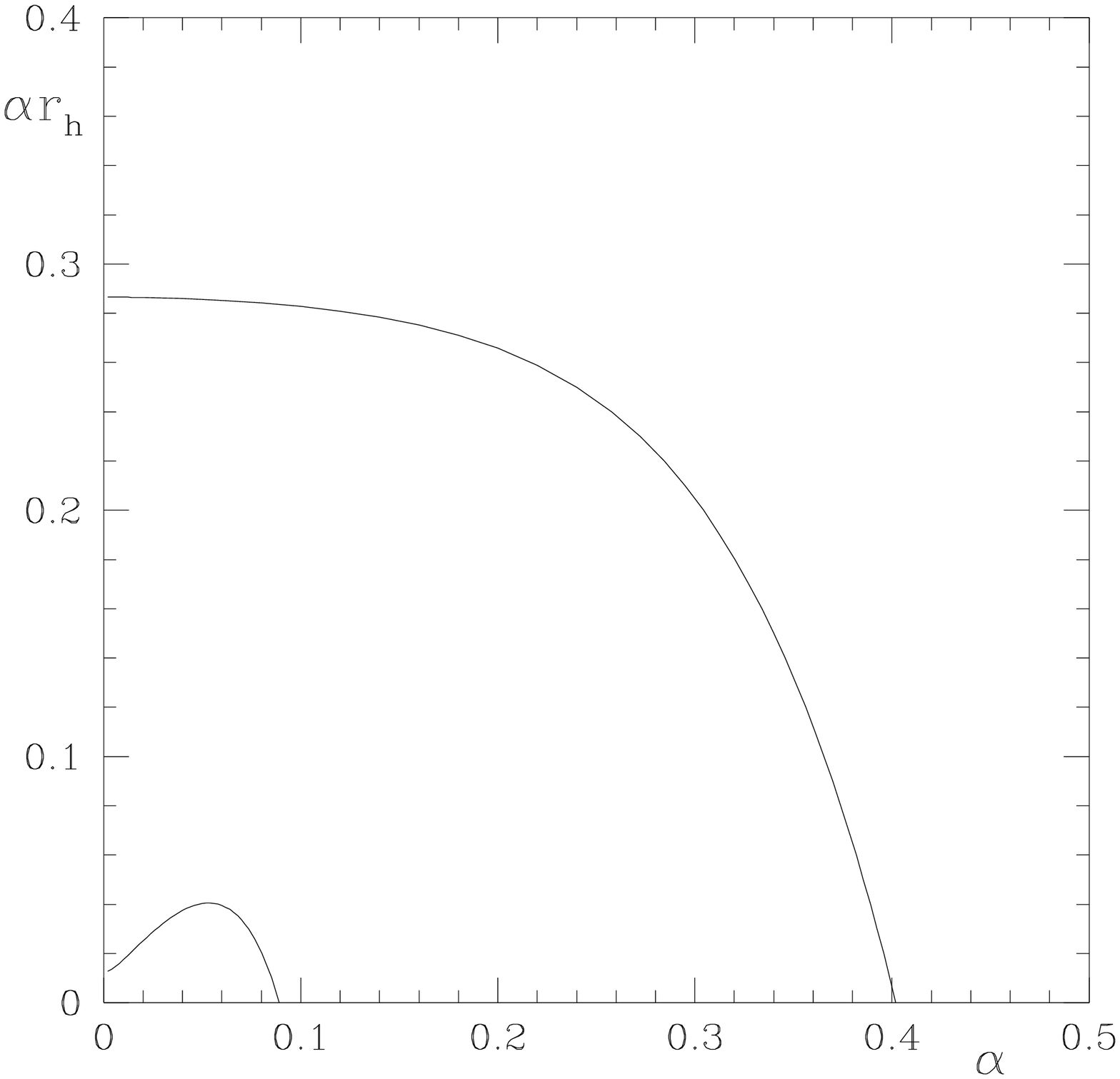,width=0.48\hsize,%
          bbllx=1.1cm,bblly=5.8cm,bburx=20.5cm,bbury=19.5cm}
}
\caption[figbalf]{\label{figbalf}
a) The parameter $b$ vs.\  $\alpha$ of the gravitating sphaleron 
 and the first two excited ones for $\beta=0$;\\
b) domains of existence for sphaleronic black holes,
fundamental sln.\ (outer boundary) and first excited sln.\ (inner boundary)
for $\beta=\infty$}
\end{figure}

Following the 1-parameter family of solutions obtained by varying $\alpha$
one finds a similar phenomenon as observed for the gravitating monopoles:
$\alpha$ runs through a maximum. 
In contrast to the situation for the
monopole there is however no critical value for $\alpha$ and the family may
be continued all the way back to $\alpha=0$.
In fact, there is no analogue of the abelian RN solution in this case,
since finite mass requires the asymptotic condition $W\to -1$ 
for $r\to\infty$.
The parameter $b(\alpha)$ (compare l.h.s.\ of Fig.~\ref{figbalf}) instead 
increases monotonously and reaches the
value $b_1$ of the BM solution from below as $\alpha$ comes back to 0. 
Thus as $\alpha$ increases the gravitational force becomes stronger and
eventually replaces the effect of the Higgs field and we end up with a
gravitationally bound sphaleron - the BM solution.
The limit $\alpha\to 0$ corresponds to case ii) discussed above.
However, Fig.~\ref{figbalf} shows that there is another branch of 
solutions starting at $b_1$ with 
$b(\alpha)\gsim b_1$ for small values of $\alpha$.      
Looking at the solution for $b\gsim b_1$ we see, that it consists of a tiny
BM solution sitting inside an essentially flat-space sphaleron of size
$1/M_W$.  
The solution starts with $W=-1$ at $r=0$, reaches almost $W=1$ within
distance $1/M_{\rm Pl}$ and then decreases slowly to $W=-1$ (compare
Fig.~\ref{figsph} r.h.s.). For all these sphaleron solutions $W$ tends to
$-1$ for $r\to\infty$ although this is not obvious from the plots, because
of the different length scales involved.
As we increase $\alpha$ it again runs
through a maximum and eventually comes back to 0 at $b=b_2$.
This whole story repeats itself, a new branch of solutions starting at each
BM solution. The values $b_n$ may be interpreted as points
where the $n^{\rm th}$ BM solution bifurcates with the same BM solution
having a flat-space sphaleron attached to it at large $r$.

Again there are solutions for $\beta=\infty$ (i.e.\ $H\equiv v$), which
one might call `global sphalerons'' in analogy to the global monopoles
of the triplet model.

Besides the globally regular solutions there are also in this case
non-abelian black holes. Like their regular counter-parts they carry no
charge. 
As already mentioned above the abelian RN solution is not allowed in this
case. 
For $\alpha\to 0$ these sphaleronic black holes tend to the 
corrersponding BM-type black holes.
Again they exist only in a bounded domain of the $\alpha$-$r_h$-plane
($r_h$ measured in units of $1/M_W$!). The domains for the various radial
excitations of the fundamental solution (with $n\geq2$ zeros of $W$) 
are nested and shrink very quickly with $n$
(compare r.h.s.\ of Fig.~\ref{figbalf}).

\section{Stability of sphalerons and sphaleronic black holes}\label{stabsph}

Like their flat counter-parts the gravitating sphalerons are unstable with
respect to certain variations involving the component $W_2$
of the YM potential \cite{Bosch}.
 
In the gravitating case we observe however an additional new type of instability 
involving variations of $W$, i.e.\ within the minimal ansatz for the YM
field.
In contrast to the first mentioned instability already present in flat space
and which we may call
``topological'', the second one may be considered a ``gravitational''
instability \cite{Lav}.
It sets in at the right turning point of the lowest branch of $b(\alpha)$,
i.e.\ at the maximal value of $\alpha$ of the gravitating version of the
flat-space sphaleron. According to a general result of stability theory
the whole upper branch between
$\alpha_{\rm max}$ and $\alpha=0$ has one unstable mode. At each turning
point of the subsequent branches of $b(\alpha)$ another unstable mode
appears. This explains the observation that the 
n$^{\rm th}$ BM solution, obtained when $\alpha\to 0$ on the 
upper part of the n$^{\rm th}$ branch, has $n$ unstable modes within the
minimal ansatz in addition to the (also $n$)
topological unstable modes involving $W_2$ \cite{Volkov}.

Analogous results are found for the sphaleronic black holes.

\end{document}